\documentclass[]{HLreport}

\usepackage{booktabs}
\usepackage{afterpage}
\usepackage[T1]{fontenc}
\usepackage{textcomp}
\usepackage{graphicx,mathptmx,amsmath,amsfonts,amscd,amsthm,amssymb,eucal,psfrag,color,subfig,url}
\usepackage{ctable}
\usepackage{mathrsfs}
\usepackage[load-configurations=abbreviations]{siunitx}
\usepackage{comment}
\usepackage[utf8]{inputenc}
\definecolor{myDarkBlue}{RGB}{0,0,192}
\usepackage[colorlinks=true,citecolor=myDarkBlue,linkcolor=myDarkBlue,linktoc=page,urlcolor=myDarkBlue]{hyperref}

\DeclareGraphicsExtensions{.pdf,.png}

\newcommand{\pT} {\mbox{$p_{\rm T}$}}

\typist{}
\documentlabel{Version 2}

\begin{document}

\setlength{\parindent}{0pt}
\setlength{\parskip}{6pt}

\title{A next-generation \\ LHC heavy-ion experiment }

\author{\it List of authors in appendix}

\abstract{
The present document discusses plans for a compact, next-generation multi-purpose detector at the LHC as a follow-up to the present ALICE experiment. The aim is to build a nearly massless barrel detector consisting of truly cylindrical layers based on curved wafer-scale ultra-thin silicon sensors with MAPS technology, featuring an unprecedented low material budget of 0.05\% X$_0$ per layer, with the innermost layers possibly positioned inside the beam pipe. In addition to superior tracking and vertexing capabilities over a wide momentum range down to a few tens of MeV/$c$, the detector will provide particle identification  via time-of-flight determination with about 20~ps resolution. In addition, electron and photon identification will be performed in a separate shower detector. The proposed detector is conceived for studies of pp, pA and AA collisions at  luminosities a factor of 20 to 50 times higher than possible with the upgraded ALICE detector, enabling a rich physics program ranging from measurements with electromagnetic probes at ultra-low transverse momenta to precision physics in the charm and beauty sector.
}

\maketitle

\tableofcontents

\newpage

\section{Introduction}
\label{sec:introduction}
%
With this document, we express our interest for a new LHC experiment, dedicated to the high-statistics study of the production of heavy flavour hadrons and of the soft electromagnetic and hadronic radiation produced in high-energy proton-proton and nuclear collisions. The apparatus would be centered on an ultra-low-mass silicon tracker, made with Complementary Metal-Oxide-Silicon (CMOS) Monolithic Active Pixel Sensors (MAPS) technology. Such an experiment could be installed during the LHC Long Shutdown 4 (LS4), at the Interaction Point 2 (IP2), where the ALICE experiment is currently installed.

The ALICE Collaboration is completing the construction of an upgraded Inner Tracking System (ITS2) with seven layers of CMOS MAPS, to be installed during LS2 \cite{ITS_upgrade}. Recent advances in this technology have made possible the fabrication of wafer-scale ultra-thin silicon detectors, and the ALICE Collaboration is currently considering a further upgrade (ITS3) in which the three innermost layers of ITS2 would be replaced with three truly cylindrical layers of such detectors, with a material budget of only 0.05\% X$_0$ per layer \cite{LS3_upgrade}.

This novel technology opens up the possibility of constructing a new, all-silicon tracker, with unprecedented low mass, that would allow reaching down to an ultra-soft region of phase space, to measure the production of very-low transverse momentum lepton pairs, photons and hadrons at the LHC. The detector would consist of a barrel and two end-caps made of layers of ultra-thin Si-sensors and cover the rapidity region $|\eta| < 4$. The ultra-low material thickness, combined with the placement of the first detector layers either inside the beam pipe or at a very close distance from its outer wall, would allow charged particle detection at transverse momenta of the order of a few tens of MeV/$c$. 

The excellent timing resolution ($\simeq 20$ ps) achievable with CMOS detectors will provide particle identification information. Electrons at low momentum ($< 500$ MeV/$c$) will be separated from hadrons using time-of-flight information while, at higher momenta, electrons and photons will be identified in a dedicated shower detector. Removable converter structures will also allow the measurement of photons with the conversion method.

Such an experiment would, for instance, allow to measure the primordial electromagnetic radiation emitted  by the Quark-Gluon Plasma (QGP) produced in nuclear collisions, providing key information for the understanding of the emergent properties of QCD matter. More generally, it would open a new window for the study of soft phenomena in hadronic collisions, allowing to address fundamental physics questions that could not be tackled so far.

The high-rate capabilities of MAPS will allow the experiment to run at significantly (a factor 20 to 50) higher luminosities than the upgraded ALICE experiment. It could therefore exploit  all pA and AA luminosities that could be reached by accelerating  ions from light to very heavy in the LHC \cite{Citron:2018lsq}.

The detector concept discussed in this document provides unprecedented physics performance for heavy flavour studies, enabling the measurement of the production of exotic quarkonia and Multiply Heavy Flavoured (MHF) baryons and mesons in pp, p-A and nuclear collisions, for which theoretical uncertainties on the cross-section span across orders of magnitude, providing a significant new window on the properties of the Quark-Gluon Plasma. 

The current ideas for a possible layout are presented in section~\ref{sec:detector_concept} and some of the areas where such an experiment would have a significant impact are discussed in section~\ref{sec:physics_potential}.

\section{Detector concept}
\label{sec:detector_concept}
%
In the following we describe the key technologies, the conceptual layout and the main features of the proposed experimental apparatus, which is entirely based on CMOS technology.
\subsection{Detector technology}
CMOS technology, which fueled the rapid growth of the information technology industry in the past 50 years, has also played and continues to play a crucial role in the remarkable development of detectors for High-Energy Physics (HEP) experiments. The amazing evolution of CMOS transistors in terms of speed, integration and cost decrease, allowed a continuous increase of density, complexity and performance of the front-end and readout circuits for HEP detectors. With the advent of CMOS MAPS, where the sensing layer and its readout circuitry are combined in a single integrated circuit, CMOS became also the technology for a new generation of vertex and tracking detectors. An important example of such an approach is represented by the new ALICE Inner Tracking System (ITS2) \cite{ITS_upgrade}, which is based on a MAPS device, named ALPIDE, covering about 10\,m$^2$ of area with about 12.5 billion pixels.

The development of ALPIDE represents a quantum leap in the field of CMOS MAPS for single-particle detection, reaching unprecedented performance in terms of signal/noise ratio, spatial resolution, material budget and readout speed. Still, further significant improvements are possible by fully exploiting the rapid progress that this technology is making in the fields of imaging and time-of-flight measurements for consumer applications. 

One of the features offered recently by CMOS imaging sensor technologies, called stitching, will allow developing a new generation of large-size MAPS with an area of up to \SI{14x14}{\cm}, for \SI{200}{\mm} wafer processes, and up to \SI{21x21}{\cm}, for \SI{300}{\mm} wafer processes. Moreover, the reduction of the sensor thickness to values of about 20\,-\,40\,$\mu$m will open the possibility of exploiting the flexible nature of silicon to implement large-area curved sensors. In this way, it will become possible to build cylindrical layers of silicon-only sensors, which will enable a dramatic reduction of the detector material thickness \cite{LS3_upgrade}. 

CMOS technology is also currently revolutionizing the field of 3D imaging, which has become a key sensing technology in a wide range of LiDAR (Light Detection And Ranging) applications in the field of robotic, automotive, medical and spacecraft systems. Time-of-flight (TOF) imagers based on CMOS Single Photon Avalanche Diodes (SPADs) feature very small pixel pitches, O(\SI{40x40}{\um}), and very high time resolution. Such a technology can be further optimized and tailored towards the measurement of minimum ionizing particles with a time resolution of the order of a few tens of picoseconds.

\subsection{Detector layout and main parameters}
\begin{figure}[h!]
\centering
\includegraphics[width=\textwidth]{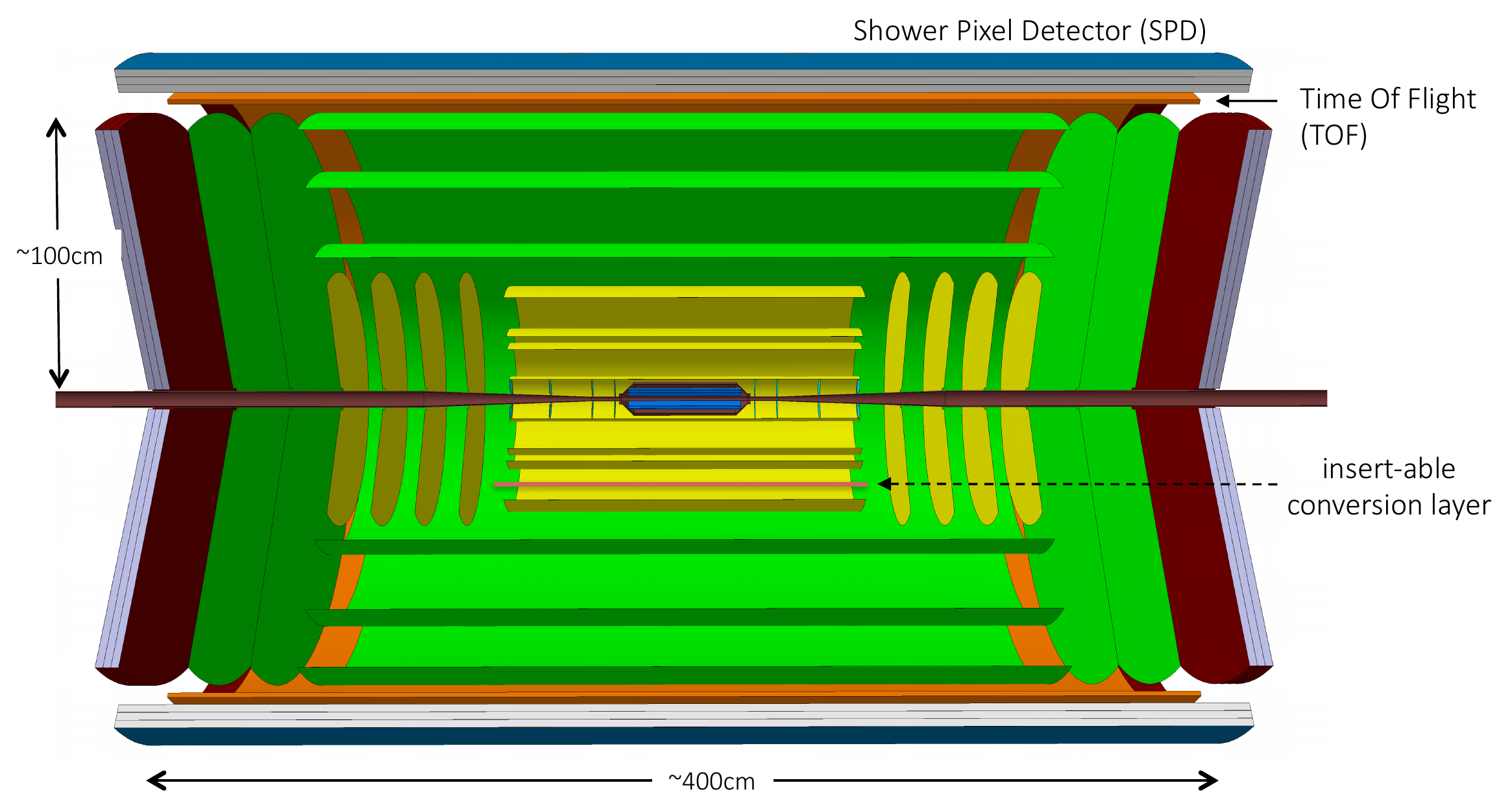}
\caption{Longitudinal view of the experimental apparatus. The detector is embedded in a solenoid magnet (not shown in the figure). The central barrel, which covers the pseudorapidity region $|\eta| < 1.4$, is composed of the IT, inside the beampipe (blue layers in the figure), the OT (yellow and green layers), the TOF (orange layers) and the SPD (outermost blue layers). Two endcaps, each consisting of a set of tracking disks and an SPD disk, extend the rapidity coverage to the region $|\eta| < 4$.}
\label{fig:detector_layout}
\end{figure}

The view of the experimental apparatus is shown in Fig.~\ref{fig:detector_layout}. The detector, which covers the pseudorapidity region of $|\eta| < 4$ over the full azimuth, has a very compact layout with radial and longitudinal dimensions of 1.2\,m and 4\,m, respectively. It consists of a central barrel and two end-caps, which are embedded in a solenoid magnet (not shown in Fig.~\ref{fig:detector_layout}). The moderately weak solenoidal field (0.5~T) of the ALICE magnet seems adequate to meet the requirements of high tracking efficiency at very low transverse momentum (a few tens of MeV$/c$) while preserving very good relative momentum resolution ($\approx 2\%$) at high transverse momentum ($\pT\approx 30~$GeV$/c$). However, the option of a magnet with a larger magnetic field (1~T or larger) will also be considered.   

The central barrel, which covers the pseudorapidity region $|\eta| < 1.4$, consists of (from the inside out) an Inner Tracker (IT), with 3 layers located inside the beam pipe, the Outer Tracker (OT) with 7 layers, a Time-Of-Flight (TOF) detector for the identification of hadrons, as well as electrons at very low transverse momentum ($\pT < 500~$MeV$/c$), and an electromagnetic Shower Pixel Detector (SPD) for the identification of electrons and photons ($\pT > 500~$MeV$/c$). The two endcaps, which extend the acceptance to the pseudorapidity region $1.4 < |\eta| < 4$, contain each 4 disks in the IT,  6 disks in the OT and one disk in the SPD.  

The IT barrel is based on curved wafer-scale ultra-thin CMOS MAPS arranged in truly cylindrical layers, featuring an unprecedented low material budget of \SI{0.05}{\percent}\,X$_0$ per layer, with the innermost layer positioned at only \SI{18}{\mm} radial distance from the beam line. The pixel sensor contains approximately $10^6$ pixels per square centimeter, each measuring about \SI{10x10}{\um}, featuring a position resolution of better than \SI{3}{\um}. The IT layers will be located either in the secondary vacuum of a larger vacuum chamber, as shown in  Fig.~\ref{fig:detector_layout}, or outside the beam pipe, with the innermost layers at a radial distance of \SI{2}{\mm} from its outer wall. Four end-cap disks on each side complement the central barrel layers. A system of seven barrel layers and six endcap disks form the OT. They are based on the same technology as the sensors for the vertex detector, but their pixel size is increased to about \SI{30x30}{\um} in order to reduce the power density. These layers will provide a spatial resolution of about \SI{5}{\um}. The material budget of the OT layers will be about \SI{0.5}{\percent}\,X$_0$ per layer. 

A TOF detector consisting of three layers of CMOS MAPS with time resolution of the order of \SI{20}{\ps} will surround the central tracker. Low-Gain Avalanche Detectors (LGAD) featuring such a time precision, which are being developed for the phase-II upgrade of the ATLAS and CMS experiments, represent a viable option. Given the moderately low radiation levels at the location of the time-of-flight system ($10^{12}$ 1\,MeV~$n_{\rm eq}$/cm$^{2}$), CMOS SPADs also represent a very promising technology. These detectors are based on arrays of avalanche photodiodes reverse-biased above their breakdown voltage. 
SPAD detectors of recent generation feature a time jitter of tens of picoseconds. The number of layers needed for the realization of the TOF detector will depend on the time resolution and spatial fill factor achieved in the single layer.

The SPD is based on a stack-up of a few layers made of a dense passive material, e.g. lead, interleaved with layers of high-granularity pixel sensors, which allow counting individual particles generated in a shower. Electrons and positrons will be distinguished from photons by the presence of a track in the tracking layers. The contamination from pions will be very small due to the large ratio between the nuclear interaction length and the radiation length of lead ($\lambda_{n}/X_{0} \simeq 30$). 

\subsection{Tracking performance}
The tracking performance of this detector has been studied using a fast Monte Carlo tool. The code, which accounts for multiple scattering, detector occupancy and deterministic energy loss, provides accurate determination of the tracking resolution as a function of the detector configuration for both the spatial and the momentum components and a reliable estimate of the tracking efficiency.

\begin{figure}
\centering
\includegraphics[width=0.8\textwidth]{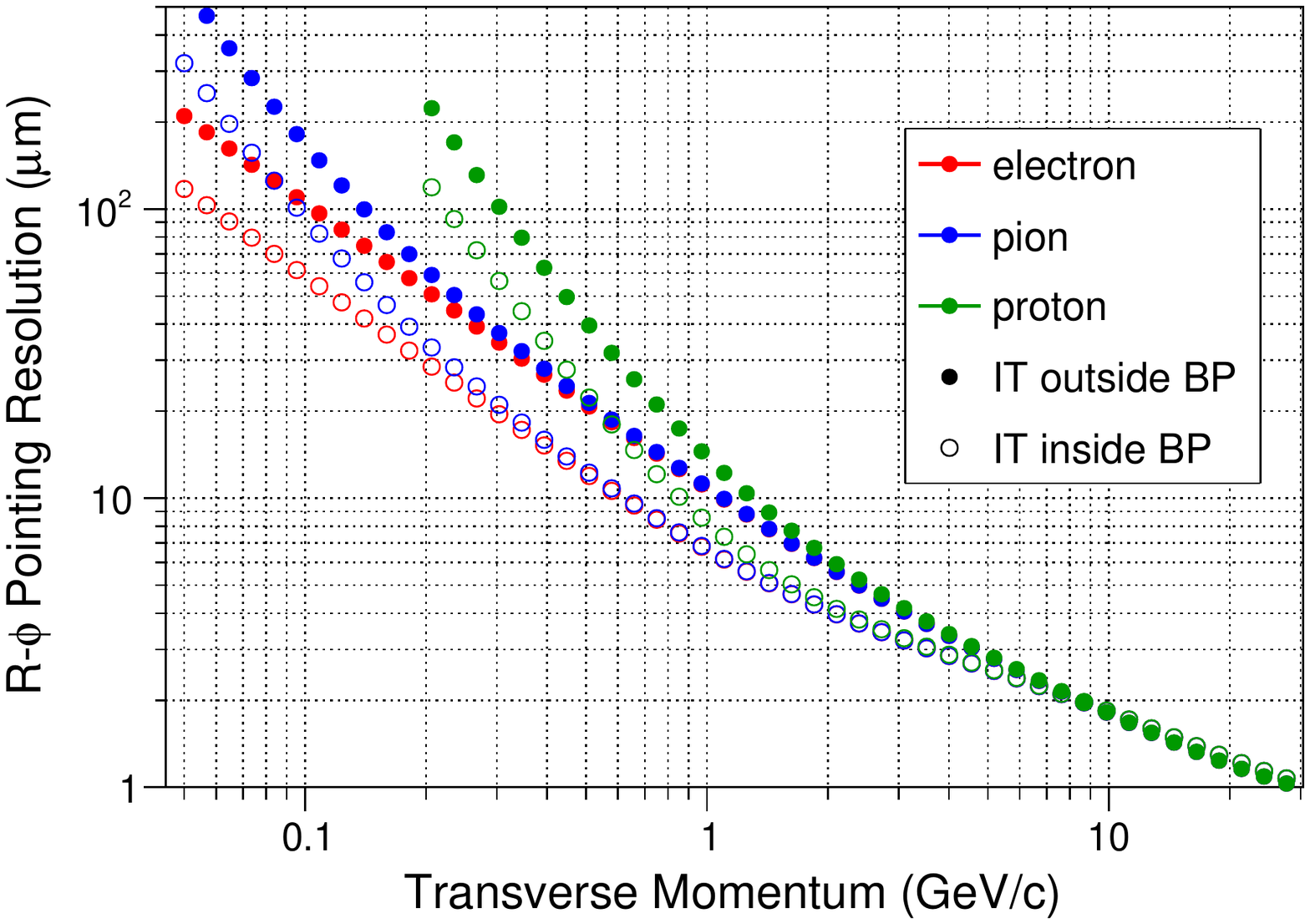}
\caption{Impact parameter resolution for primary electrons, charged pions and protons as a function of the transverse momentum for two configurations of the IT: inside the beam pipe (empty circles) and outside the beam pipe (filled circles).}
\label{fig:impact_parameter}
\end{figure}

An important measure of the achieved tracking precision is the track impact-parameter resolution, defined as the dispersion of the distribution of the Distance of Closest Approach (DCA) of the reconstructed tracks to their production vertex. It is the parameter that defines the capability of a vertex detector to separate secondary vertices of heavy-flavour decays from the interaction vertex. Two alternative configurations have been studied. In the first configuration,
the three IT layers are contained inside the beam vacuum chamber, while the innermost OT layer is located outside of it at a radial distance of about 2\,mm from the outer wall. In the second configuration the IT layers are located outside the beam pipe with the innermost layers at about 2\,mm from the outer wall. The radial arrangement of the IT layers is identical for the two configurations. 
A comparison of the impact-parameter resolution for the two configurations is shown in Fig.~\ref{fig:impact_parameter} for at typical pseudorapidity $|\eta| = 0.5$. Owing to its nearly zero mass, the detector exhibits a spectacular vertexing performance, with an impact parameter resolution that is below \SI{10}{\um} for $\pT\simeq 1~$GeV$/c$ and remains below \SI{100}{\um} down to $\pT\simeq 0.1~$GeV$/c$. 

\section{Physics potential}
\label{sec:physics_potential}
%
The ability to measure the production of leptons, photons and identified hadrons down to $\pT$ scales of the order of a few tens of MeV/$c$, would provide significant advances in several areas. In the following, we list only some of the physics topics that would dramatically benefit from such an apparatus.

\subsection{Heavy flavour and quarkonia}

The unique tracking and vertexing capabilities of the proposed apparatus, combined with its high-luminosity capabilities, will provide an ideal tool for a comprehensive campaign of high-precision measurements of the production of open and hidden heavy flavour particles in nuclear collisions. Below, we list some examples of measurements where the proposed experiment will have a crucial impact.

The suppression of the production of quarkonia due to Debye screening of the strong interaction was one of the first predicted signatures of colour deconfinement \cite{Matsui:1986dk}. 
Anomalous suppression of the production of J/$\psi$ mesons (beyond the level that could be explained by conventional models, such as the interaction of the J/$\psi$ with comoving hadrons) was experimentally observed in nuclear collisions at the CERN Super Proton Synchrotron \cite{Alessandro:2004ap}. Measurements on Au-Au collisions at RHIC revealed similar levels of suppression as at the SPS \cite{Adare:2006ns}.

Data from Pb-Pb collisions at the LHC have revealed a new regime for J/$\psi$ production~\cite{Abelev:2013ila}: at the LHC the suppression is reduced compared to lower energies and, unlike at lower energies, it is very weakly dependent on the centrality of the collision and instead of decreasing, it increases with increasing $\pT$.
Such observations are consistent with the presence of a significant contribution at the LHC from a novel mechanism, which had been proposed in~\cite{BraunMunzinger:2000px}, for a recent review see~\cite{Andronic:2017pug}, for quarkonium production via statistical hadronization of initially uncorrelated heavy quark pairs at the phase boundary. 
Alternatively, a mechanism of continuous creation and destruction of charmonia in the QGP was proposed in~\cite{Thews:2000rj}.
In this new regime for J/$\psi$ production the above mentioned $\pT$ dependence arises naturally, as predicted in~\cite{Zhao:2011cv}.
If the above picture is correct, measurements of exotic hadrons would provide a direct window on hadron formation from a deconfined Quark-Gluon Plasma, and spectacular effects would be expected for Multiply Heavy Flavoured (MHF) baryons and mesons. 
In such a scenario, the yields of MHF baryons relative to the number of produced charm quarks are predicted to be enhanced in AA relative to pp collisions. First predictions in this area were made in~\cite{Becattini:2005hb}, recent calculations are discussed in~\cite{Cho:2017dcy,Zhao:2016ccp,He:2014tga,Yao,Andronic_exotica}.  Enhancements are expected by as much as a factor 10$^2$ for the recently discovered  $\Xi_{cc}$  baryon and even by as much as a factor 10$^3$ for the as yet undiscovered $\Omega_{ccc}$ baryon. The observation and precise quantification of such effects would represent a quantum jump for the study of the properties of deconfined matter. The observation of the effect in MHF hadrons would provide a key confirmation of the current interpretation of the LHC quarkonium results
and open a crucial new window to study the mechanisms of hadron formation from a deconfined Quark-Gluon Plasma. 
Such studies are currently way beyond reach. The detection of MHF baryons at the LHC requires nucleon-nucleon luminosities of the order of 1 fb$^{-1}$ \cite{LHCb}, with excellent tracking and secondary vertex and particle identification capabilities down to low transverse momenta to enable the reconstruction of complex cascades of weak decays of heavy flavour hadrons.
The apparatus discussed in this document would provide the ideal detector for such measurements, combining ultra-low-mass tracking over a wide momentum range, particle identification and high-speed capabilities. Nucleon-nucleon luminosities in the fb$^{-1}$ range could be accumulated at the LHC in a few weeks with the use of lighter nuclei such as Xe or Kr \cite{Citron:2018lsq}, bringing such measurements well within reach.

The proposed detector would also significantly enhance the capability for quarkonium physics.
The systematics  of  charmonium  production  cannot  be  fully  understood without a precise separation of prompt charmonia from those originating from B decays as well as a quantitative determination of the production yields of low transverse momentum $\chi_c$ states, that can decay strongly into prompt J$/\psi$, but are expected to be much more fragile, due to the significantly smaller binding energy.
The separation of prompt charmonia from secondary charmonia from B decays in the proposed apparatus will be excellent, and the high luminosity should make even the rare $\psi$(2S) abundant.
The detection of $\chi_c$ states involves the identification of a low-energy (300-400 MeV) photon in addition to a J$/\psi$ meson. Low-energy photons can be measured with high efficiency in the SPD. A resolution of about 5\% around 400\,MeV is required, corresponding to $\frac{\delta E_{\gamma}}{E{\gamma}} \approx 3\%/\sqrt{E_{\gamma}(\mathrm{GeV})}$, which may be achievable in the SPD. 

The proposed detector can also be used to shed new light on the nature and structure of the X, Y, Z charmonium-like states recently discovered, see, e.g.,  \cite{Esposito:2014rxa} and references therein. As a case in point we note that the X(3872) state which is well studied in pp collisions, but the whose nature is not yet understood, could be measured with precision also in nuclear collisions. 
Since this is a charmonium state, its yield is expected to be particularly enhanced at low transverse momenta ($\pT < 4$ GeV/$c$) \cite{Andronic:2019wva}. It would also be very interesting to perform a precision comparison of the transverse momentum spectra of particles of similar mass, but very different binding energies, such as $\psi$(2S) and X(3872).

\subsection{Low-mass dileptons}

\subsubsection{The low mass continuum, $0 < m < 3$ GeV}
The strongly-interacting medium is expected to emit electromagnetic radiation during a significant part of its lifetime. At colliders, signals of an excess of  dileptons with respect to the expectations from known hadronic sources  have been observed both at RHIC~\cite{Adam:2018tdm,Adam:2018qev,Adare:2015ila} and with still low significance at the LHC~\cite{Acharya:2018nxm}. For LHC Run3 and Run4 the ALICE collaboration will upgrade its detector \cite{Abelevetal:2014cna} resulting in significant improvements for low-mass dilepton measurements compared to the results from the present RHIC and LHC experiments. 

The nearly massless detector with high-precision tracking and vertexing and very high rate capability, discussed in this document, would allow pushing down in transverse momentum very close to the natural scale determined by the inverse radius of the system (about 100 MeV/$c$ in pp collisions, about 10 MeV/$c$ in Pb-Pb collisions), which would represent the "ultimate" dilepton experiment.

The very low mass and $\pT$ cutoff would allow testing theoretical predictions in the region of phase space, currently beyond reach, where most of the radiation is emitted.  

\subsubsection{Chiral symmetry restoration and the temperature of the hot QGP fireball} 

Up to now, no direct experimental evidence exists for the restoration of chiral symmetry in the hot and dense phase 
formed during a relativistic nucleus-nucleus collision. Indeed, the enhanced low-mass dilepton continuum found at the SPS, RHIC, and LHC, can be described well by assuming that the $\rho$ meson broadens in the medium without mass change~\cite{Arnaldi:2008fw,Adamova:2006nu,Rapp:2009yu,Adare:2015ila,Acharya:2018nxm}. With an essentially mass-less detector as described above one can take an entirely new approach, based on a precision measurement of the thermal dilepton continuum starting from the $\rho$ meson and reaching up to masses of about 1.6 GeV. Guidance comes from inspecting the light flavor section of Fig. 51.3 of the latest PDG Review of Particle Physics~\cite{Tanabashi:2018oca}. Here one can see that the R factor obtained from $e^+e^-$ collisions and representing the `vacuum' in this context has a clear minimum near masses of 1.26 GeV, the mass of the a$_1$ meson. This meson is generally considered the chiral partner of the $\rho$ meson
and as such does not couple to $e^+e^-$ pairs, hence the minimum. The idea for the proposed measurement is then to study this mass region in pp, p-Pb and Pb-Pb collisions to see whether the minimum near 1.26~GeV fills in as one goes from minimum bias pp to central Pb-Pb collisions. 

With the dramatic improvements in vertexing from the  detector described above, it will be possible to quantitatively measure and tag/reject non-prompt dileptons, revealing the true thermal continuum. We note that the only other known background for this measurement is from Drell-Yan production which at LHC energies is estimated to be negligible. With this approach one could not only provide a crucial test for chiral symmetry restoration in the $\rho$-a$_1$ sector but also make a precision determination of the temperature of the QGP from the analysis of the mass spectrum in the 1.8\,--\,3 GeV region.

\subsection{Soft and ultra-soft photons} 
Measurements of real soft photons are notoriously difficult due to the huge background from $\pi^0$ decays and electron bremsstrahlung. Available measurements in ultra-relativistic nuclear collisions typically extend down to around 1 GeV, see~\cite{Adam:2015lda} and references therein. 
In the proposed ultra-low mass tracker soft real photons could be measured using the conversion method pioneered by ALICE~\cite{Adam:2015lda}. 

This should extend the photon transverse momentum range into the region of  50\,-\,100~MeV/$c$, thereby allowing tests of current predictions for radiation from the QGP in completely uncharted regions of phase space.

Very low $\pT$ photons (1~MeV/$c$ $< p_{\rm{T}}^{\gamma}<$ 100~MeV/$c$) could be measured with a special, small spectrometer at forward rapidity in the range $3.5<|\eta|<5$.

The measurement of very soft electromagnetic radiation in the p$_T$ region below 100~MeV/$c$ and approaching  1~MeV/$c$ is of fundamental interest. 
The production of photons at such low transverse momenta arises as a consequence of the structure of all gauge theories, see, in particular, \cite{bloch_nordsieck,low,cahn,strominger1,strominger2}. According to resulting soft  theorems the number of soft (real) photons actually diverges towards low $\pT$, but, as discussed in \cite{strominger2}, {\it "in a highly controlled manner that is central to the consistency of the underlying quantum field theory"}. It would be of prime importance to reach the experimental sensitivity to test this prediction. This would require measurements at very low $\pT$, below 10 MeV/$c$, which could be achieved with a high-rapidity photon spectrometer.
It is intuitively clear that the $1/p_{\rm T}$ divergence characteristic of photon bremsstrahlung and predicted by ~\cite{bloch_nordsieck,low} will eventually be reached, but at what level and at which value of $\pT$ will depend on the size and structure of the system, possibly on its quark content~\cite{nachtmann_reiter}.

To make progress in this area one would measure photon production for pp and pA and ultimately Pb-Pb collisions. Measurements performed at fixed target energies in bubble chambers~\cite{Chliapnikov:1984ed, Botterweck} and at the OMEGA Spectrometer \cite{Banerjee, Belogianni_1, Belogianni_2} reported an excess of photons at low transverse momenta (below 40 MeV/$c$) over the rate expected from hadronic bremsstrahlung, centered a positive centre-of-mass rapidity, while measurements performed with the HELIOS apparatus at central and backward rapidities~\cite{antos} reported yields consistent with the expectation from hadronic bremsstrahlung.
These results were analyzed by~\cite{botz}, in the framework of quark synchrotron radiation. It would be very important to follow up on this with a state-of-the-art measurement at the LHC.

\subsection{Other topics}
An ultra-low mass, high-resolution and high-rate experiment as discussed in this document, would also allow to make major contributions to other areas, not covered here, such as precision studies of spectral distortions at low transverse momenta, coherent pion production, hadronisation at very-low transverse momenta, disoriented chiral condensates, femtoscopy, fluctuations and diffusion of conserved charges as well as the search for dark photons.

\newpage
\bibliographystyle{ieeetr}
\bibliography{references}

\newpage
\textbf{\Large Author List}

\bigskip

D.~Adamov\'{a}$^{\rm 72}$,
G.~Aglieri Rinella$^{\rm 23}$,
M.~Agnello$^{\rm 21}$,
Z.~Ahammed$^{\rm 102}$,
D.~Aleksandrov$^{\rm 67}$,
A.~Alici$^{\rm 5,39}$,
A.~Alkin$^{\rm 2}$,
T.~Alt$^{\rm 47}$,
I.~Altsybeev$^{\rm 82}$,
D.~Andreou$^{\rm 23}$
% ...

\bigskip

\bigskip 

% ...
$^{2}$ Bogolyubov Institute for Theoretical Physics, National Academy of Sciences of Ukraine, Kiev, Ukraine\\
% ...
$^{5}$ Centro Fermi - Museo Storico della Fisica e Centro Studi e Ricerche `Enrico Fermi', Rome, Italy\\
% ...
$^{21}$ Dipartimento DISAT del Politecnico and Sezione INFN, Turin, Italy\\
% ...
$^{23}$ European Organization for Nuclear Research (CERN), Geneva, Switzerland\\
% ...
$^{39}$ INFN, Sezione di Bologna, Bologna, Italy\\
% ...
$^{47}$ Institut f\"{u}r Kernphysik, Johann Wolfgang Goethe-Universit\"{a}t Frankfurt, Frankfurt, Germany\\
% ...
$^{67}$ National Research Centre Kurchatov Institute, Moscow, Russia\\
% ...
$^{72}$ Nuclear Physics Institute of the Czech Academy of Sciences, \v{R}e\v{z} u Prahy, Czech Republic\\
% ...
$^{82}$ St. Petersburg State University, St. Petersburg, Russia\\
% ...
$^{102}$ Variable Energy Cyclotron Centre, Homi Bhabha National Institute, Kolkata, India\\
% ...

\end{document}